\documentclass[aps,pre,showpacs,twocolumn,groupedaddress,amssymb]{revtex4}

\usepackage{graphicx}
\usepackage{dcolumn}
\usepackage{bm}

\begin{document}

\title{Two stream instabilities in degenerate quantum plasmas}

\author{S. Son}
\affiliation{18 Caleb Lane, Princeton, NJ 08540}
\date{\today}

\begin{abstract}

The quantum mechanical effect on the plasma two-stream instability is studied 
based on the dielectric function approach. 
The analysis suggests that 
the degenerate plasma relevant to the inertial confinement fusion behaves 
differently from classical plasmas when 
 the electron drift velocity is comparable to  
the Fermi velocity. 
For high wave vector comparable to the Fermi wave vector, 
the degenerate quantum plasma has larger regime of the two-stream instabilities than the classical plasma. 
A regime, where the plasma waves with the frequency larger than 1.5 times of the Langmuir wave frequency become unstable to the two-stream instabilities, is identified. 
\end{abstract}

\maketitle

\section{Introduction}
Dense plasmas become  important research subjects 
 as 
plasmas with extremely high density are achieved in  
the laboratories~\cite{tabak, sonprl, sonpla} 
and as our understanding of the astrophysical entities such as 
the white dwarf and super dense stars gets  deeper. 
 The two-stream instability  in dense plamsas is especially  
important to understand 
 since 
dense  electron beams
are common phenomenon in the inertial confinement fusion (ICF)~\cite{monoelectron, tabak, ebeam} and 
the violent astrophysical events in the dense astrophysics, involving the two-stream instabilities, are also being observed from the phenomena such as the gamma ray burst~\cite{Stone, Biskamp, Fishman, Nakar, Innes}.

It becomes a challenge to understand  the dense plasmas,
as the quantum effects begin to modify the physical process at high density; 
a few physical processes 
deviating from the classical prediction  
have been identified~\cite{hass, hass2, sonprl, sonpla, sonpla2}. 
The main goal of this paper is to study the electron quantum effect on the two stream instabilities in dense quantum plasmas. 
There has been a general theoretical development in  addressing 
the quantum effects on the two-stream instabilities by employing 
the fluid-type  equation~\cite{hass, hass2}. 
In this paper, however, the author will
utilize  the more realistic Lindhard dielectric   function~\cite{Lindhard} 
and will   focus on 
the completely degenerate electron plasma  with the density 
from $n_e = 10^{24} / \mathrm{cc} $ to  $n_e = 10^{28} / \mathrm{cc} $.   
 Two cases are studied; the fist case is  
one in which the plasma has two different group of electrons with different drift velocities and the second case is  one in which 
the electrons have different velocity with the ions. 

The findings are as follows. 
When  the electron drift velocity is comparable to the Fermi velocity, 
the quantum plasma and classical plasma behave very differently. 
 For high wave vector comparable to the Fermi vector, 
the two-stream instability regime from the quantum plasmas  
is larger 
than the classical plasmas with the temperature comparable to the Fermi energy. 
 A regime is identified where 
the unstable Langmuir wave can have higher frequency than the Langmuir wave frequency by 1.5 times.

\begin{figure}
\scalebox{0.3}{
\includegraphics[width=1.7\columnwidth, angle=270]{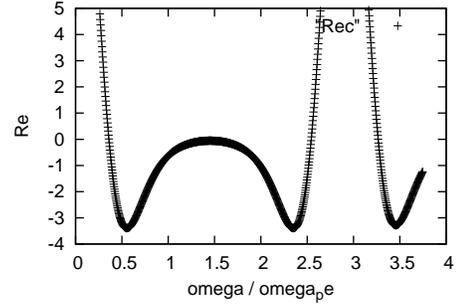}}
\caption{\label{fig1}
The real part of the dielectric function $\epsilon $ as a function of the frequency for the classical plasmas. The x-axis is $\omega / \omega_{pe} $ and the y-axis is $Re[\epsilon] $.  
In this example, $n_e = 10^{24} / \mathrm{cc} $, $T_e = 21 \  \mathrm{eV} $, $k \lambda_{de} = 0.26 $ and the drift velocity has the electron kinetic energy of 720eV so that 
 $k v_0 /\omega_{pe} \cong  2.29 $.  
The local maxima of the real part at $ \omega = 1.5  \ \omega_{pe} $ is less than 0 so taht the plasma is unstable to the two-stream instability. 
}
\end{figure} 
 
This paper is organized as follows. In Section II, the dielectric function approaches in the two-stream instability is introduced. 
In Section III, we consider the case when there are two different group of degenerate electron with different drift velocities. 
In Section IV, we consider the case when the degenerate electrons have different drift velocity with the ions. 
In Section V, the summary and discussion are provided.

\begin{figure}
\scalebox{0.3}{
\includegraphics[width=1.7\columnwidth, angle=270]{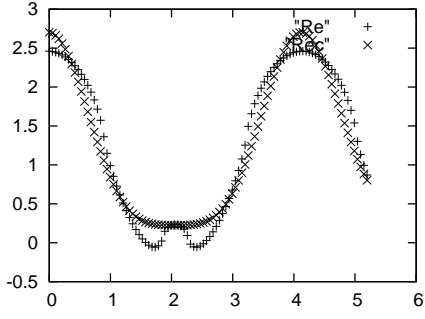}}
\caption{\label{fig2}
The real  part of the dielectric function $\epsilon $ as a function of the frequency for the completely degenerate plasma and  
the classical plasmas. 
 The x-axis is $\omega / \omega_{pe} $ and the y-axis is $Re[\epsilon] $.  
In this example, $n_e = 10^{24} / \mathrm{cc} $, $T_e = 21 \  \mathrm{eV} $, $k \lambda_{de} = 0.75 $ ($k = 0.7 k_F$) and the drift velocity has the electron kinetic energy of 320 eV so that 
 $k v_0 /\omega_{pe} \cong  4.11 $.  
The local maxima of the real part at $ \omega = 2.0  \ \omega_{pe} $ is larger than 0 and there is no instability in this wave vector.  
}
\end{figure}

\section{Dielectric Function for the two stream instability analysis.}
The longitudinal dielectric function of a plasma 
is given as   

\begin{equation} 
\epsilon(\mathbf{k}, \omega) = 1 + \frac{4 \pi e^2 }{k^2} \Sigma \chi_i \mathrm{.}
\end{equation} 
where the summation is over the group of particle species and $  \chi_i $ is the particle susceptibility. 
Given the dielectric function $\epsilon(k,\omega)$, 
the analysis of the two-stream instability can be obtained 
by finding the root of $\epsilon(k, \omega)$ as a function of $\omega$; If the dielectric function has a root $\epsilon(k, \omega) = 0$ as a function of $\omega$ in the complex upper-half plane,   the plasma is unstable to the two-stream instabilities. 
In classical plasmas, the susceptibility is given as 

\begin{equation}
 \chi_i^C(k, \omega) = \frac{n_iZ_i^2}{m_i} \int \left[ \frac{ \mathbf{k} \cdot \mathbf{\nabla}_v f_i }{\omega - \mathbf{k} \cdot \mathbf{v} }\right]d^3 \mathbf{v} 
\end{equation} 
where $m_i$ ($Z_i$, $n_i$) is the particle mass (charge, density) and $f_i $ is the distribution with the normalization $\int f_i d^3 \mathbf{v} = 1$.  
For the degenerate electrons, 
the susceptibility $\chi_e $ of the free electron plasma has been computed by Lindhard~\cite{Lindhard} and is given as

\begin{figure}
\scalebox{0.3}{
\includegraphics[width=1.7\columnwidth, angle=270]{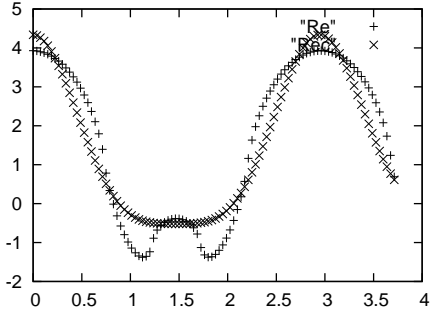}}
\caption{\label{fig3}
The real part of the dielectric function $\epsilon $ as a function of the frequency for the completely degenerate plasma and  
the classical plasmas. 
In this example, $n_e = 10^{24} / \mathrm{cc} $, $T_e = 21 \  \mathrm{eV} $, $k \lambda_{de} = 0.53 $ ($k = 0.5 k_F$) and the drift velocity has the electron kinetic energy of 320 eV so that 
 $k v_0 /\omega_{pe} \cong 2.94 $.  
The classical plasma has no local maxima. 
The local maxima of the real part at $ \omega = 1.3  \ \omega_{pe} $ for the case of the degenerate plasma  is larger than 0 and there is instability in this wave vector.  
}
\end{figure}

\begin{equation}
\chi_e^Q(\mathbf{k},\omega) = \frac{3 n_e}{m_e v_F^2} h(z, u) \mathrm{,} \nonumber \end{equation}    
where $v_F = \sqrt{2E_F/m_e}$ is the Fermi velocity, $E_F = \hbar^2 k_F^2 / 2 m_e $ ($k_F = (3 \pi^2 n_e)^{1/3}$ is the Fermi energy (Fermi wave vector), 
$ z = k / 2 k_F$, $u = \omega /k v_F$, and $h= h_r + i h_i$. 
The Fermi energy is given as
$ E_F =  36.4 \times (n/n_{24})^{2/3} \mathrm{eV}$ where  $n_{24} = 10^{24} / \mathrm{cc}$.
The real part of $h$ is given as 

\begin{eqnarray}
h_r = \frac{1}{2} + \frac{1}{8z}\left( 1- (z-u)^2\right) 
\log \left( \frac{|z-u+1|}{|z-u+1|} \right) \nonumber  \\
+ \frac{1}{8z}\left( 1- (z+u)^2\right) 
\log \left( \frac{|z+u+1|}{|z+u+1|} \right)\nonumber \\ \label{eq:re} \nonumber 
\end{eqnarray}

In the next two sections, we consider the two cases of two-stream instabilities. The first case is when there are two groups of electrons. Each group is the Maxwellian distribution (the degenerate Fermi distribution) with the same density and temperature but with different drift.
The second case is when the Maxwellian (complete degenerate) electrons have different drift velocity with the ions. The ions is assumed to be the Maxwellian with the temperature comparable to the Fermi energy. 
In the first case, the dielectric function is given as 

\begin{eqnarray} 
\epsilon =  1 &+& (4 \pi e^2/k^2) ( \chi_e^C(\omega, k) +  \chi_e^C(\omega-\mathbf{k}\cdot \mathbf{v}_0, k) \nonumber \\ \nonumber \\
 \mathrm{or} \ \  1 &+& (4 \pi e^2/k^2) ( \chi_e^Q(\omega, k) +  \chi_e^Q(\omega-\mathbf{k}\cdot \mathbf{v}_0, k) \mathrm{,} \label{eq:ele}
\end{eqnarray} 
 where  $  \mathbf{v}_0 $ is the drift velocity. 
In the second case, the dielectric function is given as
\begin{eqnarray} 
\epsilon =  1 &+& (4 \pi e^2/k^2) ( \chi_e^C(\omega, k) +  \chi_i^C(\omega-\mathbf{k}\cdot \mathbf{v}_0, k)  \nonumber \\ \nonumber \\ 
 \mathrm{or} \ \   1 &+& (4 \pi e^2/k^2) ( \chi_e^Q(\omega, k) +  \chi_i^C(\omega-\mathbf{k}\cdot \mathbf{v}_0, k)  \label{eq:ion} \mathrm{.}
\end{eqnarray}

\section{when there are two different groups of (degenerate) electrons}

\begin{figure}
\scalebox{0.3}{
\includegraphics[width=1.7\columnwidth, angle=270]{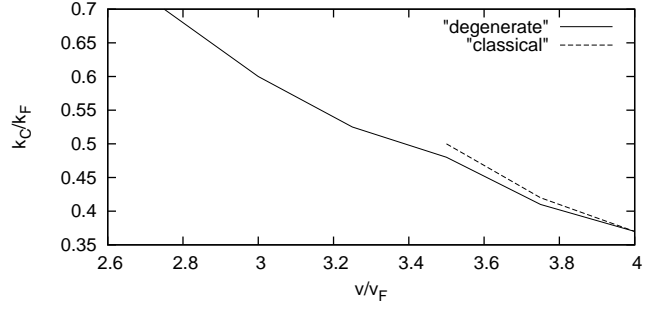}}
\caption{\label{fig4}
The boundary wave-vector $k_C $ as a function of the drift velocity. 
In this example, $n_e = 10^{24} / \mathrm{cc} $, $T_e = 21 \  \mathrm{eV} $ and $E_F = 36 \ \mathrm{eV} $. The x-axis is $v_0 / v_F $ and the y-axis is $k_c / k_F $.  
For the classical plasmas, the instability begin to emerge when $v_0 / v_F \geq 3.5 $. For the degenerate plasmas, 
the instability begin to emerge when $v_0 / v_F \geq 2.75 $.
}
\end{figure}

In this section, we analyze Eq.~(\ref{eq:ele}).  
In the conventional classical two-stream instability, the threshold condition 
for the two stream instabilities is that there should be a local maxima between $\omega_{pe} < \omega < k v_0 $ and that local maxima should be  less than 0. 
In Fig.~(\ref{fig1}), we plot the classical dielectric function 
as a function of $\omega $ for a particular $k$. 
The hump at $\omega \cong 1.3 \ \omega_{pe} $ is the local maxima and it is less than 0; the plasma Langmuir wave is unstable to  the two stream instabilities of the classical plasmas.

In Fig.~(\ref{fig2}), we plot $\epsilon$ as a function of $\omega $ for a fixed $k$ in an example plasma.  The classical plasma does not have the local maxima. The degenerate plasma has a local maxima at $\omega \cong 2 \  \omega_{pe}$ but is larger than zero.  For this particular $k$, the plasma Langmuir wave is stable to the two-stream instabilities. 
In Fig.~(\ref{fig3}), the only difference in the physical parameter with  Fig.~(\ref{fig2}) is the wave vector.  In this case, the classical plasma does not have the local maxima and the degenerate plasma has the local maxima smaller than 0;  the degenerate plasma (classical plasma) Langmuir wave is unstable (stable) to the two-stream instability.

For a fixed electron density and temperature (or completely degenerate), 
we plot $\epsilon$ as a function of $\omega $ and determine the critical $k_C(v_0)$ values for which the plasma becomes unstable to the two-stream instabilities. The  necessary condition for the two-stream instabilities  is $k < k_C(v_0)$.
By varying the drift velocity $v_0$, we can determine the boundary of the two-stream instabilities.  
In Fig.~(\ref{fig4}), we do this for a particular plasma with  $n_e = 10^{24} \  / \mathrm{cc} $, $T_e = 21 \  \mathrm{eV} $.  The Fermi energy is $E_F = 36 \ \mathrm{eV} $ and we choose the electron temperature of the classical plasma as $T_e = 0.6 \ E_F $ since  the average kinetic energy of the electron in the completely degenerate case is $0.6 \ E_F$. 
From Fig.~(\ref{fig4}), it can be concluded that 
the regime of the two-stream instability is larger in quantum prediction than  classical prediction.  As the drift velocity becomes larger than $v_0 > 3.5 \ v_F $, the difference between the quantum plasma and the classical plasma is small. 
Similar analysis suggests that, 
for more dense plasma, 
 the regime of the quantum deviation, where the degenerate plasma (classical plasma) is unstable (stable),  widen further.

\begin{figure}
\scalebox{0.3}{
\includegraphics[width=1.7\columnwidth, angle=270]{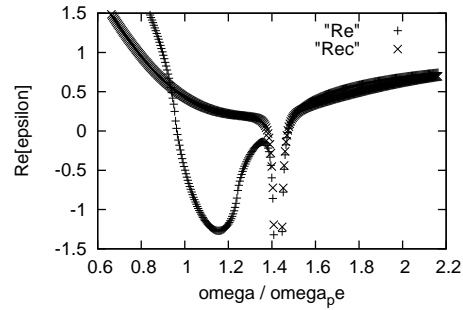}}
\caption{\label{fig5}
The real part of the dielectric function $\epsilon $ as a function of the frequency for the classical (degenerate) plasmas. 
 The x-axis is $\omega / \omega_{pe} $ and the y-axis is $Re[\epsilon] $.  
In this example, $n_e = 10^{26} / \mathrm{cc} $, $E_F = 784 \ \mathrm{eV}$,
 $T_e = 0.6 \ E_F $, $k \lambda_{de} = 0.26 k_F $ and $v_0 = 1.3  \ v_F $
so that 
 $k v_0 /\omega_{pe} \cong = 1.43 $.  
The local maxima of the real part of $\epsilon$ in the degenerate (at $ \omega = 1.4  \ \omega_{pe} $) is less than 0 and thus the plasma Langmuir  is unstable
to the two-stream instabilities.
There is no local maxima for the classical plasmas.  
}
\end{figure}

\section{when the (degenerate) electrons have different drift velocity from the ions} 
In this section, we analyze Eq.~(\ref{eq:ele}).  
We employ the same criteria that was used in Sec.~II. 
In Fig.~(\ref{fig5}), we plot the real part of the dielectric function of the classical plasmas and the degenerate plasmas. 
The condition for the two-stream instability is the existence of the local maxima of the dielectric function that is lower than 0. 
In this section, we assume that the ion temperature is zero and the ion can be treated as the classical particle as the de Broglie wave length of the ion is smaller than any other length scale in the regime of our interest.

\begin{figure}
\scalebox{0.3}{
\includegraphics[width=1.7\columnwidth, angle=270]{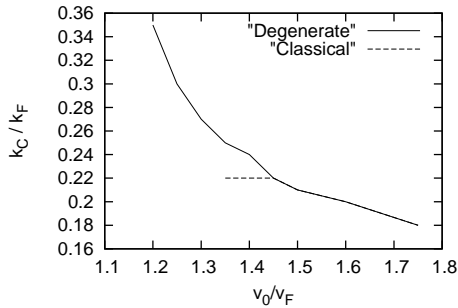}}
\caption{\label{fig6}
The boundary wave-vector $k_C $ as a function of the drift velocity. 
In this example, $n_e = 10^{26} / \mathrm{cc} $,  $E_F = 784 \ \mathrm{eV} $ and $T_e = 0.6 \ E_F$.  
 The x-axis is $\omega / \omega_{pe} $ and the y-axis is $Re[\epsilon] $.  
The x-axis is $v_0 / v_F $ and the y-axis is $k_c / k_F $.  
For the classical plasmas, the instability begin to emerge when $v_0 / v_F \geq 1.35 $. For the degenerate plasmas, 
the instability begin to emerge when $v_0 / v_F \geq 1.20 $.
}
\end{figure}

As well-known in the conventional  Lindhard function~\cite{Lindhard}, 
the degenerate plasma can support the collective Langmuir waves 
whose wave length  is much smaller than the classical Debye length. 
The regime of the two stream instability is higher in the degenerate plasma than the classical plasmas due to this fact. 
In Fig.~(\ref{fig6}), 
we plot the threshold condition 
for the  two-stream instability in the degenerate plasma and classical plasmas; 
the necessary condition is $k < k_C(v_0)$. For the classical plasmas, 
the regime of the instabilities is smaller than the degenerate plasma. 
For the plasma with the lower density, the difference between the classical plasma and the degenerate plasma is smaller than the example that we have shown.

The example in Fig.~(\ref{fig5}) is the case when 
the classical plasmas are stable regardless of the wave vector. 
As shown in the figure, the frequency of the unstable Langmuir wave is $ 1.4 \ \omega_{pe} $.  Such high frequency wave cannot be sustained in the classical plasmas and can only be supported via the diffraction and degeneracy effect of the electrons. 
For the same plasma in Fig.~(\ref{fig5}),  the unstable Langmuir wave frequency can reach $ 1.7 \  \omega_{pe} $.  

\section{summary and conclusion}
In this paper, 
the author  has analyzed the effect of the electron degeneracy and 
diffraction on the two-stream instability, 
focused on the plamsas relevant to  the inertial confinement fusion  and the astrophysical system ranging in density  
from $n_e = 10^{24} / \mathrm{cc} $ to  $n_e = 10^{28} / \mathrm{cc} $. 
Our approach is  based on the Lindhard random phase approximation, which takes into accounts the quantum degeneracy and diffraction.
We compare our prediction based on the Lindhard approaches with the classical plasma. 

The analysis suggests that
the quantum effect become more pronounced as the density gets higher and 
that the classical dielectric function describes, as accurately as the Lindhard function, the two-stream instabilities  when the drift velocity is higher than 3.5 times of the Fermi velocity. 
However, when the drift velocity is smaller than 3.5 times of the Fermi velocity, 
the quantum plasma and classical plasmas behave quite differently to the two-stream instability. 
Prominently, for the both cases we consider in Secs. II and III, 
the regime of the instability predicted by the Lindhard approach is considerably larger than the classical plasma as illustrated in Figs.~(\ref{fig4}) and (\ref{fig6}). 
We also have shown that the unstable Langmuir wave due to the two-stream instability can have the frequency as large as $1.7 \ \omega_{pe} $, which is impossible in the classical plasmas. 
Theses findings in this paper  can have major implications 
on the Backward Raman scattering~\cite{sonbackward, sonlandau,sontwo, McKinstrie, Fisch, Fisch2, sonforward, sonplasmon} and the beam stopping by the dense background plasmas.

\end{document}